\documentclass[preprint,showpacs,preprintnumbers,amsmath,amssymb]{revtex4}
                                                                                
\usepackage{graphicx}
\usepackage{psfig}
\usepackage{dcolumn}
\usepackage{bm}

\begin{document}

\title{Quantum Computation of Jones' Polynomials}
\author{V.~Subramaniam} 
\affiliation{Department of Electrical Engineering, Indian Institute of
Technology -- Bombay, Mumbai 400076, India}
\email{subbu@ee.iitb.ac.in}
\author{P.~Ramadevi}
\affiliation{Department of Physics, Indian Institute of Technology -- 
Bombay, Mumbai 400076, India}
\email{ramadevi@phy.iitb.ac.in}

\date{\today}

\begin{abstract}
It is a challenging problem to construct an efficient quantum algorithm 
which can compute the Jones' polynomial for any knot or link obtained from 
platting or capping of a $2n$-strand braid. 
We recapitulate the construction of braid-group representations from vertex 
models. We present the eigenbases and eigenvalues for the
braiding generators and its usefulness in 
direct evaluation of Jones' polynomial. The calculation suggests
that it is possible to associate a series of unitary operators
for any braid word. Hence we propose a quantum algorithm 
using these unitary operators as quantum gates acting on a $2n$ qubit state. 
We show that the quantum computation gives Jones' polynomial for 
achiral knots and links.
\end{abstract}

\pacs{03.67, 03.65}

\maketitle

\section{Introduction}
Quantum algorithms have proven to be more efficient than classical 
algorithms in solving a number of problems \cite{qcom}. 
For instance, Shor's quantum algorithm can factor numbers exponentially 
faster than classical algorithms. 
Grover's search algorithm does not change the complexity class, but 
provides significant speed up for large databases. 
We would like to explore the power of quantum algorithms in the context 
of knot theory.

Classification of knots and links in a three-dimensional space is one of
the open problems. Jones introduced a recursive 
procedure for determining a polynomial relation for these knots and links. 
Jones' polynomials do classify some knots and links \cite{jon}. 
There are other generalized polynomials which improve the classification but 
none of them have achieved complete classification \cite{lika}.
It is known that the evaluation of Jones' polynomial classically is a 
$\#{\cal P}$ hard problem \cite{jaeg}. 
Hence it will be interesting to study the computation of knot and link 
polynomials using a quantum algorithm.   

There are diverse approaches in physics to obtain polynomials for knots 
and links. 
Following Alexander's theorem, any knot can be viewed as a closure or 
capping of an $n$-strand braid. 
Therefore the polynomials for knots and links can be determined by studying
representation theory of braid groups ${\cal B}_n$. 
The common ingredient in these approaches is to find different 
representations of braid groups ${\cal B}_n$. 
We now present a brief summary of some of these approaches:

1) {\bf N-state vertex models} which are two-dimensional statistical 
mechanical models where the bonds of the square lattice carry spin $n/2$ 
representations of $SU(2)$. 
The number of possible states of spin $n/2$ is denoted by $N=n+1$. 
The properties of these models are described by the so-called $R$-matrix 
which is an $N^2 \times N^2$ matrix. 
The number of nonzero elements in the $R$-matrix for $N$-state vertex models
is given by $m=\sum_{s=1}^{N-1} 2s^2 + N^2$. 
In the literature, the vertex models are referred to either as $N$-state 
vertex model or $m$-vertex models; both are equivalent. 
For example, 2-state vertex models carry spin $1/2$ on the lattice bonds and
they are equivalently called as six-vertex models where ``six'' denotes the 
number of non-zero $R$-matrix elements.
In Ref.~\cite{Wad}, braid group representations and knot polynomials from 
the $R$-matrices of $N=2,3,4$-state vertex models were obtained.

2) {\bf Chern-Simons gauge theory} is a topological field theory which 
provides a natural framework for the study of knots and links \cite{witt}. 
The knot polynomials are given by the expectation value of Wilson loop 
observables. 
In particular, the Jones' polynomial corresponds to the Wilson loop carrying
spin $1/2$ representation in $SU(2)$ Chern-Simons theory.
Clearly, arbitrary representations of any compact gauge group $G$ can result 
in generalized polynomials \cite{rama}. 
The polynomials are in the variable $q$ which is a function of the coupling 
constant $k$ and the rank of the gauge group.
The field theoretic polynomials were obtained by exploiting the connection
between Chern-Simons theory on a three-manifold with boundary and the 
corresponding Wess-Zumino-Witten conformal field theory (WZW) on the boundary. 
The polynomials crucially depended on various representations of the 
monodromy or braiding matrices in the WZW models.

Recently, Freedman et al.\ \cite{Free1} have attempted simulation of 
topological field theories by quantum computers.
The topological quantum computation proposed in Refs.~\cite{Free1}-\cite{Free3}
is at a mathematically abstract level. 
It exploits the connection between fractional quantum Hall states and 
Chern-Simons theory at the appropriate integer coupling $k$.

3) {\bf State sum method} of obtaining bracket polynomials 
\cite{Kauff1,Kauff2}. 
In Ref.~\cite{Kauff1}, the construction of a unitary representation has been 
shown for the three-strand braid ${\cal B}_3$. 
Further within this approach, it has been shown that it is not possible for 
a quantum computer to evaluate the knot polynomial. 
However, for a specific choice of the polynomial variable, the linking number 
can be determined \cite{Kauff2}.

Our aim is to determine the Jones' polynomial for any knot or link obtained 
from braids using a quantum algorithm. For this purpose,
we need to determine matrix representation for braid generators.
We will recapitulate the construction of braid group representation
from the six-vertex model.
Then, we can determine the eigenvalues and eigenstates
for these braid generators. This exercise suggests that we can associate
product of unitary operators for any braid word. Hence the unitary
transformations, corresponding to any braid word, can be implemented 
using quantum gates. It is important to stress that the quantum
computation in this paper is crucially dependent on the mapping of 
any braid word to a product of unitary operators.

The polynomials for knots and links can be directly computed by choosing a 
suitable eigenbasis of the braiding matrices.
Essentially, a quantum algorithm will determine the probability of
finding unitarily evolved initial state ($\vert in\rangle$) 
in a final state ($\vert out\rangle$):
\begin{equation}
P_K =~ \vert \langle ~out~ \vert ~{\bf U}_B~ \vert ~ in~ \rangle \vert^2
\end{equation}
where ${\bf U}_B$ represents the series of unitary operators corresponding
to the braid word.
For the braiding matrices obtained from the six-vertex model, the above 
matrix element gives the modulus-square of the Jones' polynomial 
(up to an overall constant).  

This paper is organized as follows. 
In Section~\ref{s-braid}, we present a general method to find representation 
of braid groups using N-state vertex models. 
We discuss in detail the six-vertex model, braiding eigenvalues and eigenstates.
Using these eigenstates, we evaluate Jones' polynomial.
In Section~\ref{comp}, we present a method to perform the evaluation of the 
modulus-square of the Jones' polynomial as a quantum computation 
by considering any knot or link as 
a composition of cups, a series of braiding operations and caps. 
In the concluding section, we summarize the results obtained and discuss the 
significance of the quantum algorithm.

\section {N-state Vertex Model}
\label{s-braid}
In this section, we review the construction of braid group representations 
from N-state vertex models \cite{Wad}. 
In order to compare the eigenstates of the braiding operator with the qubit 
states, the six-vertex model (spin $1/2$ on the bonds of the square lattice) 
is relevant. 
Hence, we shall present the explicit form of the $R$-matrix and the braid 
matrix for the six-vertex model.

As mentioned in the introduction, vertex models are two-dimensional 
statistical mechanical models with the spins $n/2$ lying on the bonds of a
square lattice. 
The properties of these models are described by the $R$-matrix elements 
between edge states
$(m_1,m_2)$ and $(n_1,n_2)$: $R_{m_1 m_2}^{n_1 n_2}(u)$  
where $u$ is the spectral parameter. 
Here $m_1,m_2,n_1,n_2$ take values $n/2,n/2-1,\ldots -n/2$. 
The integrability condition of these models requires the following equations 
to be satisfied:
\begin{eqnarray} 
X_{i}(u)X_{i+1}(u+v)X_{i}(v)=& X_{i+1}(v)X_{i}(u+v)X_{i+1}(u)~,\label{YB}\\
X_i(u) X_j(v)=& X_j(v) X_i(u)~~~|i-j|>1~,\label {YB1}
\end{eqnarray}
where $X_i(u)$s are called Yang-Baxter operators and the relation (\ref{YB})
is called Yang-Baxter equation. 
The explicit form of $X_i(u)$ in terms of the $R$-matrix elements is given by
\begin{equation}
X_{i}(u)=\sum_{m_1m_2n_1n_2}R^{n_1n_2}_{m_1m_2}(u) I^{(1)} \otimes I^{(2)} 
\otimes ...E_{n_2m_1}^{i} 
\otimes E_{n_1m_2}^{i+1}\otimes...I^{(n)}~,\label{vertex}
\end{equation}
\noindent where $I^{(j)}$ is the identity acting at the $j$-th position and 
$E$ is a matrix such that $(E_{n_1m_2})_{pq}=\delta_{n_1p}\delta_{m_2q}$.
The solution to Eq.~(\ref{YB}) can be written in a compact form \cite{kaul2}:
\begin{equation}
R^{n_1n_2}_{m_1m_2}(u)=\sum_{J=0}^n
\Bigl\{ \begin{matrix}
\frac{n}{2} & \frac{n}{2} & J \cr
n_{1} & n_{2} & m \cr
\end{matrix}\Bigr\}~ \lambda_J (u)~
\Bigl\{
\begin{matrix}
{\frac{n}{2}} & {\frac{n}{2}} & J \cr
m_{2} & m_{1} & m \cr

\end{matrix}
\Bigr\} \label {diag}
\end{equation}
where the terms in parenthesis are the quantum Clebsch-Gordan coefficients 
(q-CG) \cite{pas} which are nonzero if and only if $m$ takes a value in the 
range $m=-J,-J+1,\ldots +J$ and satisfies the condition $m_1+m_2=m=n_1+n_2$. 
Here $\lambda_J(u)$ is given by
\begin{equation}
\lambda_J(u) = \prod_{s=J+1}^{j_{1}+j_{2}}\sinh(s\mu + u)
\prod_{l=|j_{1}-j_{2}|+1}^J\sinh(l\mu -u)~,
\end{equation}
\noindent where the q-CG coefficient variable $q=e^{2 \mu}$. 
These solutions are spectral parameter dependent solutions. 
The explicit form of the $R$-matrix for the six-vertex model is
\begin{center}
\begin{tabular}{ c|c c c c}
 $n_{1}n_{2}$ & $(\frac{1}{2} \;\;\frac{1}{2})$& $(\frac{1}{2}\;\frac{-1}{2})$&$(\frac{-1}{2}\;\; \frac{1}{2})$&$(\frac{-1}{2}\;\frac{-1}{2})$\\
$m_{1}m_{2}$  & & & & \\ \hline 
$(\;\frac{1}{2} \; \frac{1}{2} )$ & $sinh(\mu - u)$ & 0 & 0 & 0 \\ 
$(\;\frac{1}{2} \frac{-1}{2} )$ & 0 & $-sinh(u)$ & $e^{-u}sinh(\mu)$ & 0 \\
$(\frac{-1}{2}\; \frac{1}{2} )$ & 0 & $e^{u}sinh(\mu)$ & $-sinh(u)$ & 0 \\
$(\frac{-1}{2} \frac{-1}{2} )$ & 0 & 0 & 0 & $sinh(\mu -u)$ \\  
\end{tabular}
\end{center}

We are interested in the construction of braid group representations from
these Yang-Baxter operators $X_i(u)$ \cite{Wad}. 
The defining relations of the braid group generators are exactly same as 
Eqs.~(\ref{YB},\ref{YB1}) in the absence of the spectral parameter. 
In other words, the braid group generators $b_i \in {\cal B}_n$ (spectral
parameter independent operators) are obtained from $X_i(u)$ by taking the 
spectral parameter $u \rightarrow \infty$:
\begin{equation}
\lim_{u \to \infty}X_{i}(u)=b_{i} \label{lim}~.
\end{equation}
Let us define a quantity $\sigma_{m_1 m_2}^{n_1 n_2}$ as follows:
\begin{equation}
\lim_{u \to \infty} R_{m_1m_2}^{n_1n_2}= \sigma_{m_2 m_1}^{n_1n_2}~.
\end{equation}
The explicit form of the $\sigma^{n_1 n_2}_{m_1, m_2}$-matrix elements in 
this limit turns out to be:
\begin{center}
\begin{tabular}{ c|c c c c}
 $n_{1}n_{2}$ & $(\frac{1}{2} \;\;\frac{1}{2})$& $(\frac{1}{2}\;\frac{-1}{2})$&$(\frac{-1}{2}\;\; \frac{1}{2})$&$(\frac{-1}{2}\;\frac{-1}{2})$\\ 
$m_{1}m_{2}$  & & & & \\ \hline 
$($1/2 1/2$ )$ & $1$ & 0 & 0 & 0 \\ 
$($1/2 -1/2 $)$ & 0 & $0$ & $-q^{1 \over 2}$ & 0 \\
$($-1/2 1/2$)$ & 0 & $-q^{1 \over 2}$ & $1-q$ & 0 \\
$($-1/2 -1/2$)$ & 0 & 0 & 0 & $1$ \\  
\end{tabular}
\end{center}

The above matrix elements indicate that we can choose a basis for the 
$n$-strand braid as $\vert m_1 m_2 \ldots m_i m_{i+1} \ldots m_n \rangle$
where $m_i$'s take values $\pm {1 \over 2}$ which are the two possible 
quantum states of the spins $s_i$'s $\equiv {1 \over 2}$. 
Hence $m_i$ denotes a single qubit basis and the above basis state is an 
$n$-qubit basis state.
From Eq.~(\ref{vertex}), we can write the action of the braiding operator 
$b_i$ on such an $n$-qubit state:
\begin{equation}
b_i \vert m_1 m_2 \ldots m_i m_{i+1} \ldots m_n \rangle 
=\sum_{n_i, n_{i+1}} \sigma_{m_i m_{i+1}}^
{n_i n_{i+1}} \vert m_1 m_2 \ldots n_i n_{i+1} \ldots m_n \rangle~,
\label {nev}
\end{equation}
where $n_i,n_{i+1}$ take values $\pm {1 \over 2}$.
Clearly, this equation is not an eigenvalue equation of the braiding operator. 
We need to diagonalise the $\sigma^{n_1 n_2}_{m_1 m_2}$ matrix 
to determine the eigenbasis of the braiding operator $b_i$. 
As we did in the case of Eq.~(\ref{diag}), we can diagonalize the spectral 
parameter independent $\sigma_{m_1m_2}^{n_1n_2}$ using the quantum 
Clebsch-Gordan coefficients matrix. The corresponding eigenstates will 
be similar to the coupled states obtained 
from uncoupled states using quantum $CG$-coefficient matrix \cite{pas}:
\begin{equation}
|J_i m )= \sum_{m_i,m_{i+1}}
\Bigl\{ \begin{matrix} s_i & s_{i+1} & J_i \cr m_i & m_{i+1} & m \cr
\end{matrix} \Bigr\}\vert m_i, m_{i+1} \rangle~, \label {qcg}
\end{equation}
where $s_i,s_{i+1}$ are the spin ${1 \over 2}$ states and 
$J_i \in s_i \otimes s_{i+1} = 0,1$. 
Thus we obtain a $2$-qubit state $\vert J_i m )$ which may be an entangled
state. 

The form of the diagonal $\sigma$ matrix in the coupled basis is:
\begin{center}
\begin{tabular}{c|c c c c}
$\vert J m)$ &{$|1\;1)$} & $|1\;0)$&$|1\;-1)$& \multicolumn{1}{c}{$|0\;0)$}
\\ 
$ \vert J m )$ &&&&\\ \hline
$|1\;1)$ & $ \lambda_{1}$ & 0 & 0 & 0 \\
$|1\;0)$ & 0 & $ \lambda_{1}$ & 0 & 0 \\
$|1\;-1)$ & 0 & 0 & $ \lambda_{1}$ & 0 \\
$|0\;0)$ & 0 & 0 & 0 & $ \lambda_{0}$ \\
\end{tabular}
\end{center}
\noindent where $\lambda_1=1$ and $\lambda_0=-q$. 
These eigenvalues are equal up to an overall normalization to the eigenvalues 
of the Wess-Zumino-Witten model monodromy matrices. 

We observe that the eigenvalues of the $\sigma$ matrix on the coupled states
$\vert J m)$ depend only on $J$ and not on $m$. 
Therefore, we can suppress the $m$ dependence on the eigenstates of the 
braiding operator and equivalently write it as a tensor product state 
involving the spin ${1 \over 2}$ placed on the bonds of the six-vertex model.
That is,
\begin{equation}
\vert J_i m) \equiv \vert J_i \rangle = \vert (s_i \otimes
s_{i+1})_{J_i} \rangle~. \label {notat}
\end{equation}
 
Even though we have explicitly diagonalized the $\sigma$ matrix, we must 
remember that all the braid group generators $b_i$'s cannot be simultaneously 
diagonalized.
The spectral parameter independent form of Eqs. (\ref{YB},\ref{YB1}) are 
the defining relations of the braid group ${\cal B}_n$ which implies that we 
can simultaneously diagonalize either $b_{2i}$'s or $b_{2i+1}$'s.  

\subsection {Representation Theory of Braid Groups in Knot Theory}
In this subsection, we would like to address the eigenvectors and eigenvalues 
of braid generators from the viewpoint of obtaining polynomial invariants 
of knots from platting or capping of braids. 

It is well known that knots from braids are not unique. 
That is, braids related by Markov moves I and II give rise to the same knot.
These two moves indeed completely remove the non-uniqueness.
So the construction of polynomial invariants for knots must be such that 
the polynomial does not change under Markov moves. 
One such procedure for knots obtained from closure of braids has been 
presented in \cite{Wad}. 

We will use the eigenstates of the braiding operators to directly compute 
the polynomial invariant for any knot obtained from closure or capping of 
an $n$-strand braid. 
In order to remove the non-uniqueness due to Markov moves, we place 
orientations on the strands of the braid. 
Further, we introduce a correction factor to the braid eigenvalues obtained 
from six-vertex model such that the polynomial does not change under Markov 
moves I and II. 
The correction factor on braiding eigenvalues depends on the relative 
orientations between the two strands. 
For right-handed half-twists between strands of parallel orientation, the 
braiding eigenvalues $\lambda_J^{(+)}$ are 
\begin{equation}
\lambda_0^{(+)} = -q^{3/2} ~~;~~\lambda_1^{(+)}= q^{1/2}~. \label {para}
\end{equation}
Similarly, for right-handed half twists between strands of antiparallel 
orientation, the braiding eigenvalues $\lambda_J^{(-)}$ are
\begin{equation}
\lambda_0^{(-)} = 1 ~~;~~\lambda_1^{(-)}=-q^{-1}~. \label {antip}
\end{equation}
The eigenvalues for left-handed half-twists are inverse of the right-handed 
half-twists eigenvalues. 

Suppose we consider any knot obtained from capping of a $2n$-strand oriented 
braid. 
Clearly, capping is possible if the number of outgoing strands is equal the 
number of incoming strands in the oriented braid. 
In other words, the quantum states $m_i$'s on the strands (\ref{nev}) should 
be such that 
\begin{equation}
\sum_{i=1}^{2n} m_i =0 ~. \label {zer}
\end{equation}
Therefore to study knots from braids, only the subspace of the states 
$\vert m_1 m_2 \ldots m_{2n} \rangle$ satisfying Eq.~(\ref{zer}) needs to be 
considered. 
Hence the construction of the eigenstates of braiding operators should be 
consistent with Eq.~(\ref{zer}). 

We will now present the eigenstates of the braiding matrices which will 
enable direct evaluation of knot polynomials.
For a $2n$-strand oriented braid, we can write the most general eigenbasis 
of braiding operators $b_{2i+1}$ with eigenvalue $\lambda_{j_{2i+1}}$, for 
all $i$'s, as:
\begin{equation}
\begin{split}
\vert \phi_{(\{J_{2i+1}\},\{l_i\})} \rangle=&\vert (~\ldots (\ldots(~(s_1 \otimes s_2)_{J_1}\otimes (s_3 \otimes s_4)_{J_3}~)_{l_1}\\ 
&\otimes \ldots(s_{2i+1} \otimes s_{2i+2})_{J_ {2i+1}})_{l_i}  
\otimes \ldots (s_{2n-1} \otimes s_{2n})_{J_{2n-1}})_0 \rangle \label {odd}
\end{split}
\end{equation}
Recall that the appropriate braiding eigenvalues (\ref{para},\ref{antip}) 
need to be substituted depending on the relative orientations of the two 
strands involved in braiding and the handedness.
The brackets within the basis kets should be identified with the notation in 
Eq.~(\ref{notat}). 
That is, $\vert (s_1\otimes s_2)_{J_1} \rangle =\vert J_1 \rangle$,
$\vert ((s_1 \otimes s_2)_{J_1} \otimes (s_3 \otimes s_4)_{J_3})_{l_1}\rangle
= \vert (J_1 \otimes J_3)_{l_1} \rangle$, and so on.
Note that the final combined state in the above basis is chosen to be spin 
$0$ which is essential to satisfy the condition (\ref{zer}) to describe 
knots from closure or capping of braids.

In the similar fashion, we can write a different eigenbasis for braiding 
operators $b_{2i}$ with eigenvalue $\lambda_{J_{2i}}$ for all $i$'s:
\begin{equation}
\begin{split}
\vert {\tilde \phi}_{(\{J_{2i}\},\{r_i\})} \rangle=&
\vert( (\ldots~(~\ldots (~(~s_1 \otimes (s_2 \otimes s_3))_{J_2})_{r_1}\otimes \ldots\\
&(s_{2i} \otimes s_{2i+1})_{J_{2i}})_{r_{i-1}}\otimes \ldots 
(s_{2n-2} \otimes s_{2n-1})_{J_{2n-2}})_{r_{n-2}}
\otimes s_{2n})_0 \rangle~\label {even}
\end{split}
\end{equation}
In order to achieve the final spin $0$ state, we require $r_{n-2}={1 \over 2}$.
Incidentally, these two bases are equivalent to the conformal blocks in 
Wess-Witten conformal field theory.

The two different bases (\ref{odd},\ref{even}) are related by an orthogonal 
(unitary) duality matrix $a_{(\{J_{2i+1}\},\{l_i\}),(\{J_{2i}\},\{r_i\})}$
\begin{equation}
\vert \phi_{(\{J_{2i+1}\},\{l_i\})} \rangle~=~a_{(\{J_{2i+1}\},\{l_i\}), 
(\{J_{2i}\},\{r_i\})}~ \vert {\tilde \phi}_{(\{J_{2i}\},\{r_i\})} \rangle~.
\end{equation}
The duality matrix can be written in terms of products of $SU(2)_q$ 
quantum-Racah coefficient matrices \cite{kal}: 
\begin{eqnarray}
a_{(\{J_{2i+1}\},\{l_i\}), (\{J_{2i}\},\{r_i\})}\nonumber 
&=&\sum_{t_1,\ldots t_{n-2}}
\prod_{i=1}^{n-2}\left(a_{t_i J_{2i+1}}\left[\begin{matrix}l_{i-1} & s_{2i+1}\cr
s_{2i+2} &l_i \end{matrix}\right] \prod_{m=0}^{n-2}a_{t_i r_{i-1}}\left[\begin{matrix}t_{i-1} &J_{2i}\cr
r_i&s_{2m}\end{matrix}\right]\right)~\nonumber\\
&~& \prod_{m=0}^{n-2} a_{l_i J_{2i+2}} \left[
\begin{matrix}t_m & s_{2m+2}\cr
s_{2m+3}& t_{m+1}\end{matrix}\right]
\end{eqnarray}
where the closed form expression for the quantum-Racah coefficient matrix 
is \cite{Res}
\begin{widetext}
\begin{eqnarray}
a_{jl}\left[\begin{matrix}s_1 & s_2 \cr
s_3 & s_4\end{matrix} \right]&=& (-1)^{s_1+s_2+s_3+s_4} \sqrt{[2j+1][2l+1]}
\Delta(s_1,s_2,j) \Delta(s_3,s_4,j) \Delta(s_1,s_4,l) \Delta(s_2,
s_3,l)\nonumber\\
~&~&\times \sum_{m \geq 0}(-1)^m [m+1]!
\{[m-s_1-s_2-j]! [m-s_3-s_4-j]! [m-s_1-s_4-l]!\nonumber\\
~&~&\times [m-s_2-s_3-l]![s_1+s_2+s_3+s_4-m]!
 [s_1+s_3+j+l-m]!\nonumber\\
~&~&\times[s_2+s_4+j+l-m]!\}^{-1}
\end{eqnarray}
\end{widetext}
where $\Delta(a,b,c)= \sqrt{[-a+b+c]![a-b+c]![a+b-c]!\over[a+b+c+1]!}$ and 
the number in square brackets represents the $q$-number defined as
$$[x]= {q^{x/2}-q^{-x/2} \over q^{1/2} -q^{-1/2}}.$$

With this detailed background on the representation theory of braid groups, 
we are now in a position to evaluate Jones' polynomial for any knot
or link.

\subsection{Evaluation of Jones' polynomial}
Consider any knot or link as shown in Fig.~1 which is technically
called platting or capping of braids.
\begin{figure}[ht!]
\centerline{\psfig{file=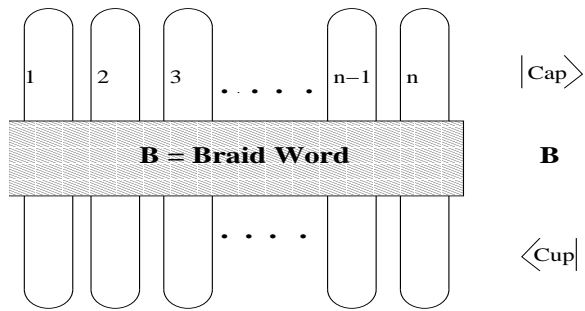,height=3in,width=3.0in}}
\caption{Viewing Knots as a Capping of Braiding Sequences} \label{graph3}
\end{figure}
The knot type is determined by the braiding region denoted as the shaded 
region in Fig.~1. 
It involves a sequence of braiding operations which is usually written as 
a braid word with appropriate orientations. 
Taking the initial state and final state as 
$\vert \phi_{(\{0\},\{0\})}\rangle$, the following
matrix element
\begin{equation}
V_K= \langle \phi_{(\{0\},\{0\})}\vert {\bf B} 
\vert \phi_{(\{0\},\{0\})} \rangle \label {jone}
\end{equation}
gives the Jones' polynomial (up to an overall
normalisation). To see the polynomial form in variable $q$, we 
can substitute the eigenvalues of the braiding generators
and the unitary duality matrices elaborated in the previous
subsection. 

For the variable $q$ being root of unity, we can equivalently 
denote the sequence of braiding operations ${\bf B} \in {\cal B}_{2n}$ 
as a product of unitary $2^{2n} \times 2^{2n}$ matrices.
We will now illustrate this mapping for two examples involving four-strand
and six-strand braids.  
As shown in Fig.2(a), the braid word
$$B_a= b_2^3 {\hat b_1}^{-2} {\hat b_3}^{-2} b_2^3 \in {\cal B}_4~,$$
where ${\hat b_i}$ denotes braiding between 
anti-parallely oriented $i$-th strand and $i$+1-th strand
and $b_j$ denotes braiding between parallely oriented $j$-th strand
and $j$+1-th strand.
Following the rules of the representation theory of 
braids, the braid-word in Fig.~2(a) will be mapped to a product
of $2^4 \times 2^4$ unitary matrices as follows:
\begin{equation}
B_a \equiv {\bf U}_{B_a}=
\left(\{{\bf a}~ {\bf f(b_2)}~{\bf a}^{\dagger}~ {\bf g(b_1,b_3)} 
~{\bf a}~ {\bf h(b_2)} ~{\bf a}^{\dagger}\}_{2 \times 2} \oplus 
{\bf 1}_{14 \times 14}\right)_{2^4 \times 2^4}
\end{equation}
where ${\bf a}$ is the duality matrix and the 
matrix elements of  the matrices ${\bf f(b_2)}$, ${\bf g(b_1,b_3)}$  
and ${\bf h(b_2)}$, for the braid word $B_a$, will be 
\begin{equation}
{\bf f(b_2)}_{J_1 J_2}= \delta_{J_1 J_2}
(\lambda_{J_2}^{(+)})^3~,~~
{\bf g(b_1,b_3)}_{J_1,J_2}=
\delta_{J_1 J_2} (\lambda_{J_1}^{(-)})^{-4}~,~~ 
{\bf h(b_2)}_{J_1 J_2}= \delta_{J_1 J_2} (\lambda_{J_2}^{(+)})^3~.
\end{equation}
The entries in these diagonal  matrices are
functions of the braiding eigenvalues. The relative
orientations and the number of crossings between the
respective braids are incorporated in writing the functional 
form. If $q$ is  a root of unity, these diagonal matrices are 
unitary. 
Thus we see that the braid word $B_a \in {\cal B}_4$ can be 
equivalently represented as a product of unitary $2^4 
\times 2^4$ matrices. 
\begin{figure}
\centerline{\psfig{file=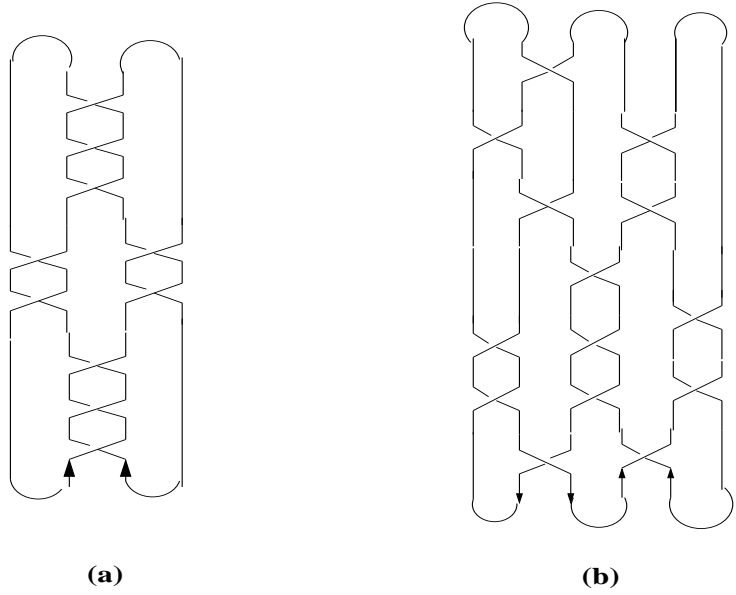,height=5in,width=5.0in}}
\caption{Two examples (a) $B_a \in {\cal B}_4$~~, (b) $B_b \in {\cal B}_6$ }
\label {graph2}
\end{figure}
Similarly, the braid word $B_b$ corresponding to Fig.~2(b) will be
$$B_b=(b_2^{-1} b_4) ({\hat b_1}^{-2} {\hat b_3}^{-3} {\hat b_5}^{-2})
({\hat b_2} {\hat b_4}^2) b_1 {\hat b_2}~ \in {\cal B}_6 ~.$$
The equivalent description involving product of $2^6 \times 2^6$ unitary
matrices will be 
\begin{equation}
{\bf U}_{B_b}= [\left({\bf a}~ {\bf f(b_2,b_4)}~ 
{\bf a}^{\dagger}~ {\bf g(b_1,b_3,b_5)}
~{\bf a}~ {\bf h(b_2,b_4)}~ {\bf a}^{\dagger}~ {\bf f_1(b_1)}
~{\bf a} ~{\bf g_1(b_2)} ~{\bf a}^{\dagger}\right)_{5 \times 5} \oplus
{\bf 1}_{59 \times 59}]_{2^6 \times 2^6}
\end{equation}
where the duality matrix elements are denoted as $a_{(J_1,J_3,J_5=l_1), 
(J_2, J_4,r_1)}$ and the remaining $5 \times 5$ diagonal matrices 
for the braid word $B_b$ will be
\begin{eqnarray}
{\bf f (b_2,b_4)}_{(J_2^{\prime},J_4^{\prime},r_1^{\prime}), (J_2,J_4,r_1)}&=&
 \delta_{J_2^{\prime} J_2} \delta_{J_4^{\prime},J_4} \delta_{r_1^{\prime}, r_1}
(\lambda_{J_2}^{(+)})^{-1} \lambda_{J_4}^{(+)} \nonumber\\
{\bf g(b_1,b_3,b_5)}_{(J_1^{\prime},J_3^{\prime},J_5^{\prime}), (J_1,J_3,J_5)}
&=& \delta_{J_1^{\prime} J_1} \delta_{J_3^{\prime},J_3} 
\delta_{J_5^{\prime},J_5}(\lambda_{J_1}^{(-)})^{-2} (\lambda_{J_3}^{(-)})^{-3}
(\lambda_{J_5}^{(-)})^{-2} 
\nonumber\\
{\bf h(b_2,b_4)}_{(J_2^{\prime},J_4^{\prime},r_1^{\prime}), (J_2,J_4,r_1)}&=&
 \delta_{J_2^{\prime} J_2} \delta_{J_4^{\prime},J_4} \delta_{r_1^{\prime}, r_1}
\lambda_{J_2^{\prime}}^{(-)} (\lambda_{J_4^{\prime}}^{(-)})^2  \nonumber\\
{\bf f_1(b_1)}_{(J_1^{\prime},J_3^{\prime},J_5^{\prime}), (J_1,J_3,J_5)}
&=& \delta_{J_1^{\prime} J_1} \delta_{J_3^{\prime},J_3} 
\delta_{J_5^{\prime},J_5}\lambda_{J_1^{\prime}}^{(+)} \nonumber\\ 
{\bf g_1(b_2)}_{(J_2^{\prime},J_4^{\prime},r_1^{\prime}), (J_2,J_4,r_1)}&=&
 \delta_{J_2^{\prime} J_2} \delta_{J_4^{\prime},J_4} \delta_{r_1^{\prime}, r_1}
\lambda_{J_2^{\prime} }^{(-)} ~. 
\end{eqnarray} 
In a similar fashion, we can find the unitary representation
${\bf U}_{B}$ for any braid word 
$B \in {\cal B}_{2n}$ in terms of products of 
$2^{2n} \times 2^{2n}$ unitary matrices
involving duality matrices 
$[a_{(\{J_{2i+1}\},\{l_i\}), (\{J_{2i}\},\{r_i\})} \oplus \bf 1]_{2^{2n}
\times 2^{2n}}$ and diagonal braiding matrices. 
These unitary representations play the role of quantum
gates in the  quantum computation of Jones' polynomial.

\section{Quantum Computation}
\label{comp}
In this section, we attempt to compute the Jones' polynomials, 
for knots and links obtained from platting or capping of
$2n$ strand braid as shown in Fig.~1,
through a quantum algorithm. We have already elaborated in
the previous section that we can associate ${\bf U}_B$ 
(product of unitary matrices) for every braid word
$B \in {\cal B}_{2n}$. The quantum algorithm 
involves the following steps:\\
\underline{Step 1}: Let the initial $2n$-qubit state be $\vert 0\rangle$ 
($\vert cup \rangle$).\\
\underline{Step 2}: We perform the sequence of unitary operations
${\bf U}_B$ corresponding to the braid word $B \in {\cal B}_{2n}$ in
Fig.~1. The unitarily transformed state will be
\begin{equation}
\vert \Psi \rangle = {\bf U}_B \vert 0 \rangle~.
\end{equation}
\underline{Step 3}: Finally, we determine the probablity of the 
unitarily evolved state $\vert \Psi \rangle$ in a specific final state 
$\vert f \rangle$ as 
\begin{equation}
\vert \langle f \vert 
{\bf U}_B \vert 0 \rangle \vert^2~.
\end{equation}
Taking the final state to be $\vert cap \rangle = \vert 0 \rangle$,
we get the modulus square of the Jones' polynomial 
(up to an overall normalisation)$V_K$ (\ref {jone}). 

For a subclass of knots(links) called achiral knots(links), 
$V_K$ is unchanged under $q \rightarrow q^{-1}$.
In other words, the matrix element $\langle 0 \vert U_B \vert 0 \rangle$
will be real. For these achiral knots and links, the quantum algorithm
directly gives the Jones' polynomial (up to an overall normalisation).

\section{Summary and Discussions}
In this paper, we have presented matrix representations for braiding 
matrices from six-vertex models. We have discussed the 
representation theory of braids, namely, the eigenbasis 
and eigenvalues of the braid generators obtained from six-vertex models.
The explicit evaluation of Jones' polynomial, for any knot/link
from braids, is presented. From the evaluation, we have
shown that we can associate a series of unitary operators
for any braid word. This is the significant result of the paper
enabling quantum computation. We have demonstrated a quantum algorithm, 
involving these unitary operators, which can determine the modulus 
square of the Jones' polynomial for any knot or link. The algorithm 
gives Jones' polynomial for achiral knots and links. 

We must realize that the quantum computation essentially determines the 
probablity of unitarily evolved initial state in a specific final state.
Further, the number of unitary operators is dependent on the
braid word and at most equal to twice the length of the braid word.

{\bf Acknowledgments}: We would like to thank L.H.~Kauffman for comments and 
queries which significantly helped us to improve the paper. 
We would also like to thank Umasankar for going over the manuscript and 
suggesting corrections. 




\end{document}